\pdfoutput=1
\documentclass[11pt]{article}

\usepackage[final]{nlp4MusA}
\usepackage{times}
\usepackage{latexsym}
\usepackage[T1]{fontenc}
\usepackage[utf8]{inputenc}
\usepackage{microtype}
\usepackage{inconsolata}

\usepackage{graphicx}
\usepackage{tabularx}  
\usepackage{lipsum,lineno}
\usepackage[most]{tcolorbox}
\DeclareUnicodeCharacter{2212}{-}
\usepackage{float}
\usepackage{caption}

\title{Beyond Musical Descriptors: \\Extracting Preference-Bearing Intent in Music Queries}

\author{
  \textbf{Marion Baranes\textsuperscript{1}},
  \textbf{Romain Hennequin\textsuperscript{1}} and
  \textbf{Elena V. Epure\textsuperscript{1, 2}}
\\    
\texttt{research@deezer.com}
\\ \\
  \textsuperscript{1}Deezer Research, Paris, France,\\
  \textsuperscript{2}Idiap Research Institute, Martigny, Switzerland.
\\ \\
}

\begin{document}
\maketitle
\begin{abstract}
Although annotated music descriptor datasets for user queries are increasingly common, few consider the user’s intent behind these descriptors, which is essential for effectively meeting their needs. We introduce MusicRecoIntent, a manually annotated corpus of 2,291 Reddit music requests, labeling musical descriptors across seven categories with positive, negative, or referential preference-bearing roles.
We then investigate how reliably large language models (LLMs) can extract these music descriptors, finding that they do capture explicit descriptors but struggle with context-dependent ones. This work can further serve as a benchmark for fine-grained modeling of user intent and for gaining insights into improving LLM-based music understanding systems.
\end{abstract}

\section{Introduction}
\label{sec:intro}
Users increasingly expect machines to understand complex and subjective natural-language queries for music search or recommendation \citep{Doh2023, Gupta2023,porcaro2019, Delcluze2025, Palumbo25, Melchiorre2025}. 
While search engines reliably handle focused queries—those naming a specific artist, track, or album—they remain much less effective for open queries where user intent is exploratory \citep{Hosey2019, Sguerra2022}. 
Addressing open queries requires interpreting complex musical descriptors like genre (e.g., \textit{pop, jazz}), mood (e.g., \textit{sad, calm}), listening context (e.g., \textit{party, driving}), instrumentation (e.g., \textit{guitar, piano}), time period (e.g., \textit{80s}), or geographical origin (e.g., \textit{Spanish music}).

However, understanding descriptors alone is insufficient: it is also necessary to determine how each relates to the user’s intent—whether it expresses a desired attribute (e.g., \textit{I want to listen some Rock}), an undesired one (e.g., \textit{Recommend me anything except Elvis Presley.}), or serves as a reference point (e.g., similarity: \textit{I want more recent music, but like Elvis Presley.}). 
We capture this distinction by assigning each descriptor a \textbf{preference-bearing intent}, expressing either \textit{positive} affinity (+), \textit{negative} aversion (-), or a \textit{referential} role with softer notions of similarity ($\sim$).
Although several works \cite{hachmeier2025, salganik2025musicsem_ismir} focus on music descriptor extraction or annotation, none consider the associated preference-bearing intents, and it remains unclear whether LLMs can robustly capture both the breadth of musical attributes and the preference-bearing roles of these descriptors.

To fill these gaps, we propose:
(1) \textit{MusicRecoIntent}, an annotated corpus of music-related queries, with each descriptor linked to a preference-bearing role, enabling fine-grained analysis of user intents.\footnote{Dataset is available at \url{https://github.com/deezer/MusicRecoIntent-NLP4MusA26}.}
(2) A benchmark of popular LLMs for this extraction task.
(3) A qualitative analysis of systematic errors in manual and automatic annotations.

\section{Related Work}
Several datasets focus on music metadata extraction, particularly named entities (NE) such as artists or track titles. \citet{Epure2022} created \textit{MusicRecoNER}, a Reddit-based corpus of music recommendation queries, showing that extracting musical entities is non-trivial. \citet{hachmeier2024, hachmeier2025} extended this work using LLMs on queries extracted from Reddit, but also YouTube.
Other studies address semantic aspects of musical descriptions. \citet{salganik2025musicsem_ismir, salganik2025_NeurIPS} introduced MusicSem, a large language–audio corpus annotated across five semantic categories, while \citet{Weck2024_WikiMuTe} extract descriptors such as genre, style, mood, instrumentation, and tempo from Wikipedia. Additional datasets, like MusicCaps \citep{agostinelli2023} or JamendoMaxCaps \citep{roy2025}, provide captions and aspect lists for musical segments, linking audio tracks with semantic descriptions.

Information Extraction (IE) aims to extract structured information from unstructured text, encompassing tasks such as Named Entity Recognition. 
In the musical domain, \citet{hachmeier2025} show that LLMs outperform smaller models in detecting musical entities, although performance strongly depends on prior exposure of LLMs to the entities. Similarly, \citet{salganik2025_NeurIPS} demonstrate that LLMs can extract detailed semantic information from music descriptions, highlighting their growing adoption in music-related IE.

Beyond IE, determining the preference-bearing role of each descriptors remains challenging. Negation detection studies indicate that LLMs often struggle, where semantically contradictory statements are treated as equivalent \citep{kim2025semantic, vrabcová2025}. Similarity detection in user musical queries has been shown by \citet{Palumbo25} to be reliably handled by LLMs.

Despite recent advances, existing datasets and solutions fail to capture complex user intentions, including negation, or softer notions of similarity. 
This work fills these gaps by introducing a dataset for benchmarking LLMs' ability to model musical descriptors and their preference-bearing roles, supporting a richer understanding of user intent.

\section{MusicRecoIntent Dataset}
Our corpus is based on \textit{MusicRecoNER} \citep{Epure2023}, which contains English-language music recommendation requests collected from Reddit\footnote{\url{https://www.reddit.com/}}. 
These requests, wrote by users for other users, have not been corrected or standardized, which explains their variable quality. 
They are mostly open-ended with their length varying according to the level of detail and complexity.
For the purpose of the present study, only one third of the dataset was retained, namely 2,291 user queries.

\paragraph{Manual Annotation.}
The annotation task aimed to label all musical descriptors in each query according to whether the user wanted them, wanted something similar, or wanted to avoid them. Two annotators were instructed to annotate descriptors in the categories introduced in Section \ref{sec:intro}, together with their corresponding preference-bearing intent.

\paragraph{Validation.}
We measured inter-annotator agreement separately for descriptor extraction and preference-bearing roles. 
Cohen’s Kappa is unsuitable for descriptor identification, which is multi-label and span-based. 
Thus, agreement was computed at the span level as a percentage. This metric naturally accounts for multiple descriptors per query and variations in annotated elements. 
For the preference-bearing intents, we relied on Cohen’s Kappa. Results are reported in Table~\ref{tab:agreement_scores}.

\begin{table}[h!]
\small
\centering
\begin{tabular}{lcrr}
\hline   & {Aggr.} & {Pref. intent}        & \# Descriptors\\
        & {(\%)}& {(Kappa)} & in Common \\
\hline
Decade      & 86.7 & 0.889 &   92 \\
Genre       & 78.2 & 0.634 &  716 \\
Instrument  & 66.1 & 1.000 &  253 \\
LC          & 66.8 & 1.000 &  160 \\
Mood        & 69.4 & 0.898 &  410 \\
NE          & 85.4 & 0.752 & 1677 \\
Country     & 81.7 & 1.000 &   53 \\
\hline
Global    & 77.1 & 0.927 & 3361 \\
\hline
\end{tabular}
\caption{\small Inter-annotators Agreements per Category}
\label{tab:agreement_scores}
\end{table}

Overall, the agreement on descriptor extraction is substantial, with a global rate of 0.771. Agreement varies across categories, ranging from 0.661 for \textit{Instrument} to 0.867 for \textit{Decade}.  The main disagreements are often due to typographical variations (e.g., \textit{oppressive / oprressive}), spontaneous normalization or correction (e.g., \textit{1980s / 80s}), segmentation differences (e.g., \textit{frank oceans blonde / frank ocean, blonde}), or context-driven additions or omissions (e.g., \textit{klassische / klassische musik}). The annotators noted the absence of a category dedicated to more structural musical information, such as rhythm or song composition.

The preference-bearing intent agreement was computed on descriptors extracted by both annotators, and is generally high: 0.927, indicating a strong agreement. 
Category-level Kappa scores range from 0.634 for \textit{Genre} to 1 for \textit{Instrument}, \textit{Listening Context} (LC), and \textit{Country}. 
Most descriptors were marked as desired (positive affinity), while named entities were frequently marked as referential, rather than being explicit targets. 
Negative preferences are quite rare across all categories.

\paragraph{Final Dataset Overview.} 
To create the final dataset, the annotators reviewed all points of disagreement together and reached a consensus on each annotation.
The final corpus contains a total of 3,935 annotations. 
Table~\ref{tab:corpus_stats} provides a detailed overview of descriptors per category, indicating whether they were annotated with a positive (+), negative (-), or referential (\textasciitilde) preference-bearing role. 
On average, each query contains between 1 and 3 elements per category.

\begin{table}[h!]
\centering
\small
\begin{tabular}{lccccc}
\hline & \#Descr. & \textbf{+} & \textbf{\textasciitilde} & \textbf{-} & Coverage\\
\hline
Decade             & 94   & 84   & 9    & 1   & 3.03 \\
Genre              & 792  & 749  & 25   & 18  & 3.37 \\
Instru.         & 352  & 337  & 1    & 14  & 1.81 \\
LC   & 217  & 215  & 2    & 0   & 1.12 \\
Mood               & 551  & 539  & 0    & 12  & 2.13 \\
NE                 & 1870 & 253  & 1613 & 4   & 1.10 \\
Country            & 59   & 57   & 2    & 0   & 1.59 \\
\hline
Total     & 3935 & 2234 & 1652 & 49  & 1.51 \\
\hline
\end{tabular}
\caption{\small Descriptive Statistics of the Corpus by Category}
\label{tab:corpus_stats}
\end{table}

\section{Automatic Annotation with LLMs}
To extract descriptors and preference-bearing intents, we rely on Ollama\footnote{\url{https://ollama.com/}}, an open-source framework that facilitates the use of a wide range of LLMs. 
We experiment with a diverse set of models and sizes: Gemma 3 (4B, 12B, 27B) \citep{Gemma2025}, LLaMA 3 (1.8B) \citep{llama2024}, Mistral (7B) \citep{Mistral7b}, and Qwen 3 (8B, 32B) \citep{Qwen2025}.
We design a single prompt that instructs the model to identify music entities, as well as the other descriptor types, using a broad definition to avoid imposing too many restrictions.  The full prompt is provided in Appendix~\ref{sec:appendix_PromptA}. To guide the model and enhance consistency and accuracy, the task was illustrated with six concrete examples. A second prompt was employed to determine the preference-bearing intent of each descriptor. This subsequent prompt is provided in Appendix~\ref{sec:appendix_PromptB}.

Table~\ref{tab:allmodel} summarizes the number of descriptors extracted by the tested models. 
Models generally produced more descriptors than those in the manually annotated dataset. 
To ensure reliability and limit hallucinations, only descriptors similar enough to the original text were retained.
The preference-bearing intent was predicted afterward for the descriptors generated by Gemma3:27b, the model with the best performance on the initial task.

\paragraph{Evaluation Metrics.}
The most common evaluation metrics used in the extraction of musical entities are precision, recall, and F-score. 
We extend these metrics to better analyze the types of errors our system may produce, following the methodology proposed by \citet{Batista2018}.
For each prediction, we classify it as correct, missing, spurious, incorrect, or partial (overlaps partially with the expected entity). Then, results are evaluated for exact match only (a prediction is considered correct if its segmentation is exact) and partial match (A prediction is considered correct if its segmentation is at least partially accurate, i.e. if it has at least one word in common with the expected prediction). It should be noted that these scores depend both on annotation quality and on model performance.

\begin{table}[h!]
\centering
\small
\begin{tabular}{lrrr}
\hline
LLM &  Exact &  Partial & \# Descr. \\
\hline
Gemma3:4b & 0.66 & 0.74 & 4405\\
Gemma3:12b & 0.68 & 0.75 & 5010\\
Gemma3:27b & \textbf{0.69} & \textbf{0.76} & 4860\\
LLaMA3.1:8b & 0.60 & 0.68 & 4535\\
Mistral:7b  & 0.56& 0.64 & 4813\\
Qwen3:8b & \textbf{0.69} & \textbf{0.76} & 4628\\
Qwen3:32b & 0.67 & 0.73 & 5034\\\hline 
\end{tabular}
\caption{\small Overall F1-scores (Exact and Partial) and Number of Descriptors Extracted by Different LLMs.}\label{tab:allmodel}
\end{table}

\paragraph{Overall Results.}
Table~\ref{tab:allmodel} shows the performance of various LLMs.
Gemma (12b, 27b) and Qwen 8b achieve the strongest overall performance, with the highest exact and partial F1-scores. 
In contrast, Llama 8b and Mistral 7b perform less consistently across descriptor categories. 
Overall, larger or more recent architectures tend to generalize better at this extraction task. 
Based on this, we selected Gemma 27B (which was faster than Qwen3-8B) for a detailed analysis and further experiments. 

\paragraph{Results on Descriptors Extraction.}
\begin{table}[h!]
\centering
\small
\begin{tabular}{l|ccc|ccc}
\hline 
\textbf{Descriptor} & \multicolumn{3}{c|}{\textbf{Exact Match}} & \multicolumn{3}{c}{\textbf{Partial Match}} \\
\textbf{categories} & P & R & F1 & P & R & F1 \\
\hline 
NE          & 0.86 & 0.82 & 0.84 & 0.93 & 0.88 & 0.90 \\
Genre       & 0.83 & 0.80 & 0.82 & 0.92 & 0.88 & 0.90 \\
Mood        & 0.85 & 0.78 & 0.82 & 0.93 & 0.85 & 0.89 \\
LC          & 0.47 & 0.42 & 0.45 & 0.74 & 0.65 & 0.69 \\
Instrument  & 0.69 & 0.62 & 0.66 & 0.85 & 0.76 & 0.80 \\
Country     & 0.96 & 0.88 & 0.92 & 0.98 & 0.90 & 0.94 \\
Decade      & 0.87 & 0.83 & 0.85 & 0.93 & 0.89 & 0.91 \\
\hline 
Overall     & 0.62 & 0.77 & 0.69 & 0.69 & 0.85 & 0.76 \\
\hline 
\end{tabular}
\caption{\small Precision, Recall and F1-scores with Gemma3:27B}
\label{tab:musical_entities_scores}
\end{table}
Tables~\ref{tab:musical_entities_scores} shows that NEs are well recognized, with an Exact Match F1-score of 0.84. 
The model accurately identifies explicit and well-defined descriptors, such as Country, which achieves a F1-score above 0.90. 
Decade, Genre, and Mood are also well extracted.
The comparison between Exact Match and Partial Match illustrates the effect of segmentation tolerance, boosting F1-score from 0.69 to 0.76.
This indicates that the model often identifies descriptors correctly, but token boundaries are imperfect. 

The relatively lower scores for Listening Context (LC) and Instrument are partly due to limitations in the manual annotation process. 
LC often involves detailed descriptors, leading to annotation variability (e.g., "\textit{I want music reco, in the summer my dad and me are doing a trip to norway}" - \textit{[summer, norway]} vs. \textit{['dad', 'summer', 'trip to norway']}), while Instrument performance is limited by segmentation issues (e.g., \textit{electric guitar} vs. \textit{guitar}). These factors explain the gap between Exact and Partial Match scores, highlighting the difficulty of achieving consistent span annotations, whether human- or LLM-generated.

\paragraph{Results on Preference-Bearing Intent Prediction.}
For this evaluation, we consider only the descriptors annotated both manually and by Gemma3:27B (3,030 descriptors).
Among these descriptors, 89\% of LLM predictions matched the ground-truth.
Table \ref{tab:confusion_matrix_gemma27} summarizes the results.
Overall, the confusion matrix shows that the model performs well, though it tends to overpredict positive preferences at the expense of referential cases---231 cases are incorrectly predicted as positive.
Although negative preferences are accurately extracted, their low frequency limits the robustness of the conclusions.

%

\begin{table}[h!]
\centering
\small
\setlength{\tabcolsep}{10pt}
\renewcommand{\arraystretch}{1.2}
\begin{tabular}{c|ccc|c}
\hline
True $\backslash$ Pred & \textbf{+} & \textbf{--} & \textbf{\textasciitilde} & \#Descr. \\
\hline
\textbf{+}                & 0.94 & 0.01 & 0.05 & 1627\\
\textbf{--}               & 0.00 & 0.90 & 0.10 & 31\\
\textbf{\textasciitilde}  & 0.17 & 0.01 & 0.82 & 1372\\
\hline
\end{tabular}
\caption{\small Normalized Confusion Matrix (per True Class) for Preference-Bearing Intent Prediction with Gemma 3:27B.}
\label{tab:confusion_matrix_gemma27_normalized}\label{tab:confusion_matrix_gemma27}
\end{table}

\paragraph{Qualitative analysis.}
When comparing manual annotations with Gemma-3 predictions (cf. Table~\ref{example}), several recurring sources of disagreement emerge.
In many cases, both the human annotator and the model identify the same underlying information but disagree on segmentation—some merge multiple tokens into a single descriptor, others split them into separate units. 
Boundary disagreements also arise from differences in expected granularity: predicted spans may be shorter or longer than the ground-truth, often reflecting annotation guideline ambiguities rather than genuine model errors.

Additional discrepancies involve truncated or slightly altered descriptors, particularly for named entities. 
These cases generally reflect minor lexical drifts rather than semantic misunderstanding. 
The model also tends to over-annotate highly generic musical terms (e.g., \textit{music, song, album}), which, while domain-relevant, do not serve as meaningful descriptors. 
Omissions are also common, especially for song titles and artist names, whose surface forms often resemble ordinary text, making them difficult to distinguish from non-descriptive content.
The scores in Table~\ref{tab:musical_entities_scores} illustrate the impact of these discrepancies. Exact match penalizes any deviation—including segmentation differences—whereas partial match tolerates boundary variations. 
Segmentation disagreements are pronounced for Listening Context and Instrument descriptors, partly due to guideline limitations that did not anticipate certain edge cases, explaining some MusicRecoIntent boundary inconsistencies.

For the second annotation task—determining the preference-bearing role of a descriptor as positive, negative, or referential—additional sources of disagreement arise.
The most frequent occur when similarity requests are interpreted as positive preferences, particularly when phrasing includes "\textit{like}" questioning the prompt design. Interestingly, nearly 80\% of these cases involve named entities. 
Less common but more challenging are sentences with strong negation (e.g., \textit{hate, not, unless}) while expressing an overall positive intent; in these cases, half of the errors concern musical genres. Examples include: "\textit{I hate most rap but want to get more into the genre}" or "\textit{Don't listen to a lot of EDM but I really like Porter Robinson’s Shelter.}"

Finally, a subset of disagreements stems from genuine ambiguity, where multiple interpretations are possible. For example, "\textit{more songs like 1000 Rounds by Pouya and Ghostemane}" may refer either to the song alone or to the song and its artists, and both readings are plausible.

\section{Conclusion}
By introducing \textit{MusicRecoIntent}, a corpus annotated with musical descriptors and preference-bearing roles, this work provides a benchmark for fine-grained modeling of user intent in music-related queries. 
Our results show that LLMs reliably capture explicit, well-defined descriptors—such as named entities, country, genre, and mood, and predict preference-bearing roles with high accuracy for positive and negative cases, while referential roles are more challenging and often overpredicted.
Beyond model performance, our analysis reveals shared challenges for both annotators and LLMs, particularly around boundary decisions, granularity, and semantic ambiguity, suggesting that future improvements will require clearer annotation guidelines and prompts.

\bibliography{redditDataset}

\appendix
\section{Disagreements Examples in the Dataset}
Table~\ref{example} shows examples of disagreements observed between manual annotations and Gemma3:27b annotations.

\begin{table}[htbp]
  \centering
  \small
  \begin{tabular}{llll}
    \hline
    \textbf{Query} & \textbf{Manual} & \textbf{Gemma3:27b} & \textbf{Disagreements}\\
        & \textbf{annotations} & \textbf{annotations} & \textbf{Type}\\
    \hline
    songs similar to japanese ceremonial &['japanese ceremonial&['japanese', 'ceremonial',&Segmentation\\
    tea &tea']&'tea']&\\ \hline
    more genre busting like wu tang vs &['wu tang vs beatles']&['wu tang', 'beatles']&Segmentation\\
     beatles &&&\\ \hline
    songs like kendrick lamars maad city&['kendrick lamars', &['kendrick lamars maad city']&Segmentation\\ 
    &'maad city']&& \\ \hline
    nostalgic indie pop alt rock songs &['nostalgic', 'indie', 'pop', & ['nostalgic', 'indie pop',&Segmentation\\ 
    &'alt rock'] & 'alt rock']& \\ \hline
    looking for a wedding first dance song&['wedding first dance song']&['wedding', 'dance']&Segmentation\\ \hline
    music similar to these songs do you &['do you feel it', 'chaos &['chaos chaos']&Omission\\
    feel it chaos chaos &chaos']&&\\ \hline
    looking for lovechild of metallica & ['lovechild', 'metallica'] & ['metallica']&Omission\\\hline
    songs based off of eminems &['eminems', 'phenomenal']&['eminem', 'phenomenal']& Truncation\\
    phenomenal&&&\\\hline
    similar to otherside by the red hot&['the red hot chili peppers', &['red hot chili peppers',& Truncation\\
      chili peppers &'otherside']&otherside']&\\\hline
    music genre for grandiose stylistic &['grandiose stylistic&['music', 'grandiose', & Over-detection\\
    trumpet & trumpet']&'stylistic', 'trumpet']&  + Segmentation \\\hline
    a sad song&['sad']&['sad', 'song']& Over-detection \\\hline
  \end{tabular}
  \caption{\small Examples of disagreements between manual annotations and Gemma3:27b annotations}\label{example} 
\end{table}

\section{Prompts}
\label{sec:appendix_Prompt}
\subsection{Prompt for Musical Descriptors Extraction}\label{sec:appendix_PromptA}
\begin{tcolorbox}[breakable,colframe=black,colback=white, boxrule=0.5pt,left=6pt, right=6pt, top=6pt, bottom=6pt] \small
You're an assistant specialized in music. Your aim is to detect and extract musical descriptors mentioned in sentences. We want to extract all types of musical descriptors cited in the given text: artist names (or group/band names) and all musical work of art as song title or album title, musical genre, decade, location, mood, listening context, instruments, etc. 
Output must be a list that can contains all the descriptors found. Here some examples of the output expected:\\
- eg 1 : text : "I love rock and roll" - output: ['rock and roll'], \\
- eg 2 : text : "I love sad spanish love songs " - output: ['sad', 'spanish', 'love']. \\
- eg 3 : text : "je veux du gros rap des années 80 pour faire la fête" - output: ['rap', 'années 80', 'fête'], \\
- eg 4 : text : "I want some french songs like j'irai ou tu iras - Céline dion and JJ Goldman"  - output: [french, "Céline dion", "JJ Goldman", "j'irai ou tu iras"].\\
- eg 5 : text : "Pop music to cook and sing" - output: ['pop', 'cook', 'sing].\\
- eg 6 : text : "I love Whenever, Wherever - Shakira " - output: ["Shakira", "Whenever, Wherever"].\\
Do not explain what you are doing, do not add any information that is not in the text to process and do not modify or correct the extracted text. Write only the output in one line. If you don't find descriptors, write [].</Task>\\
Please extract the musical descriptors cited in the following text : "{{sentence}}"
\end{tcolorbox}

\newpage

\subsection{Prompt for Preference-Bearing Intent Extraction}\label{sec:appendix_PromptB}
\begin{tcolorbox}[breakable,colframe=black,colback=white, boxrule=0.5pt,left=6pt, right=6pt, top=6pt, bottom=6pt] \small
You're an assistant specialized in natural langage processing. You will receive: a user query, and a list of descriptors that appear in the query. For each descriptor, determine the user's intention toward it using the following labels: \\
"'+' : the user is explicitly looking for this descriptor.\\
"'\textasciitilde' : the user is looking for something similar or related, but not necessarily exactly this descriptor. \\
"'-' : the user wants to exclude this descriptor (indicated by negations such as no, not, without, avoid, exclude, etc.).\\
Output format is strict: Return only a Python list of tuples of the form: [(descriptor, intention), ...]. No explanations, no extra text. \\
Rules: \\
a) Assign '+' if the descriptor is explicitly mentioned in the query as something the user wants, without being negated or rejected.\\
b) Assign '-' if the query explicitly negates, excludes, rejects (e.g., no X, not X, without X, exclude X, avoid X, etc.). \\
c) Assign '\textasciitilde' if the descriptor is mentioned in a way that suggests the user is looking for something related, similar, or loosely connected, but not exactly that descriptor.\\
Here some examples of the output expected:\\
- eg 1: User query: 'dark 90s music' ; Descriptors: '['dark', '90s']' -> Expected output: [('dark', '+'), ('90s', '+')]\\
- eg 2: User query: 'music like Abba but more rock' ; Descriptors: '['Abba', 'rock']' -> Expected output: [('Abba', '\textasciitilde'), ('rock', '+')]\\
- eg 3: User query: 'rock music without guitar' ; Descriptors: '['rock', 'guitar']' -> Expected output: [('rock', '+'), ('guitar', '-')]\\
- eg 4: User query: 'Celine dion song without Goldman' ; Descriptors: '['Celine dion', 'Goldman']' -> Expected output: [('Celine dion', '+'), ('Goldman', '-')]\\
- eg 5: User query: 'Calm rock song similar to the beatles' ; Descriptors: '['Calm ', 'rock', the beatles]' -> Expected output: [('Calm', '+'), ('rock', '+'), ('the beatles', '\textasciitilde')]\\
Now process the following instance: User query: \"{{sentence}}\" ; Descriptors: "{{desc}}"
\end{tcolorbox}

\end{document}